\documentclass[11pt]{article}

\usepackage{times}
\usepackage{sidecap}
\usepackage{geometry}
\usepackage{floatrow}
\usepackage{float}
\usepackage{hyperref}
\usepackage{graphics,graphicx}
\usepackage{epsf,wrapfig}

\usepackage[super,sort&compress]{natbib} 
\usepackage{paralist} 
\renewenvironment{thebibliography}[1]{%
\textsc{\textbf{References:}} 
\let\par\relax\let\newblock\relax%
\inparaitem[{[}1{]}]}{\endinparaitem} 

\usepackage{times} 
\usepackage{etoolbox} 
\makeatletter 
\patchcmd{\@lbibitem}{\item[\hfil\NAT@anchor{#2}{\NAT@num}]}{\item[\NAT@anchor{#2}{\NAT@num}]}{}{} 
\makeatother

\DeclareGraphicsExtensions{.jpg,.pdf,.png,.eps}

\usepackage{caption}
\usepackage{xcolor}
\usepackage{tcolorbox}

\definecolor{darkcyan}{rgb}{0.0, 0.0, 0.7}

\definecolor{boxcolor}{RGB}{193, 229, 255}
\definecolor{figboxcolor}{RGB}{193, 229, 255}
\definecolor{boxcolorgray}{RGB}{227, 227, 227}
\definecolor{darkcyan}{rgb}{0.0, 0.0, 0.7}

\def\gtrsim{\lower 2pt \hbox{$\, \buildrel {\scriptstyle >}\over
{\scriptstyle \sim}\,$}}
\def\lesssim{\lower 2pt \hbox{$\, \buildrel {\scriptstyle <}\over
{\scriptstyle \sim}\,$}}

\def\approxlt{\lower.2em\hbox{$\buildrel < \over \sim$}}
\def\approxgt{\lower.2em\hbox{$\buildrel > \over \sim$}}

\def\be{\begin{equation}}
\def\ee{\end{equation}}
\def\bea{\begin{eqnarray*}}
\def\eea{\end{eqnarray*}}

\def\nii{[N{\sc ii}]}
\def\sii{[S{\sc ii}]}

\def\oii{[O{\sc ii}]}
\def\oiii{[O{\sc iii}]}

\def\jref@jnl#1{{\rm#1}}
\def\aj{\jref@jnl{AJ}}			
\def\araa{\jref@jnl{ARA\&A}}		
\def\apj{\jref@jnl{ApJ}}			
\def\apjl{\jref@jnl{ApJ}}		
\def\apjs{\jref@jnl{ApJS}}		
\def\ao{\jref@jnl{Appl.~Opt.}}		
\def\apss{\jref@jnl{Ap\&SS}}		
\def\aap{\jref@jnl{A\&A}}		
\def\aapr{\jref@jnl{A\&A~Rev.}}		
\def\aaps{\jref@jnl{A\&AS}}		
\def\azh{\jref@jnl{AZh}}			
\def\baas{\jref@jnl{BAAS}}		
\def\jrasc{\jref@jnl{JRASC}}		
\def\memras{\jref@jnl{MmRAS}}		
\def\mnras{\jref@jnl{MNRAS}}		
\def\pra{\jref@jnl{Phys.~Rev.~A}}	
\def\prb{\jref@jnl{Phys.~Rev.~B}}	
\def\prc{\jref@jnl{Phys.~Rev.~C}}	
\def\prd{\jref@jnl{Phys.~Rev.~D}}	
\def\pre{\jref@jnl{Phys.~Rev.~E}}	
\def\prl{\jref@jnl{Phys.~Rev.~Lett.}}	
\def\pasp{\jref@jnl{PASP}}		
\def\pasj{\jref@jnl{PASJ}}		
\def\qjras{\jref@jnl{QJRAS}}		
\def\skytel{\jref@jnl{S\&T}}		
\def\solphys{\jref@jnl{Sol.~Phys.}}	
\def\sovast{\jref@jnl{Soviet~Ast.}}	
\def\ssr{\jref@jnl{Space~Sci.~Rev.}}	
\def\zap{\jref@jnl{ZAp}}			
\def\nat{\jref@jnl{Nature}}		
\def\iaucirc{\jref@jnl{IAU~Circ.}}
\def\aplett{\jref@jnl{Astrophys.~Lett.}}
\def\apspr{\jref@jnl{Astrophys.~Space~Phys.~Res.}}
\def\bain{\jref@jnl{Bull.~Astron.~Inst.~Netherlands}}
\def\fcp{\jref@jnl{Fund.~Cosmic~Phys.}}
\def\gca{\jref@jnl{Geochim.~Cosmochim.~Acta}}
\def\grl{\jref@jnl{Geophys.~Res.~Lett.}}
\def\jcp{\jref@jnl{J.~Chem.~Phys.}}	
\def\jgr{\jref@jnl{J.~Geophys.~Res.}}	
\def\jqsrt{\jref@jnl{J.~Quant.~Spec.~Radiat.~Transf.}}
\def\memsai{\jref@jnl{Mem.~Soc.~Astron.~Italiana}}
\def\nphysa{\jref@jnl{Nucl.~Phys.~A}}
\def\physrep{\jref@jnl{Phys.~Rep.}}
\def\physscr{\jref@jnl{Phys.~Scr}}
\def\planss{\jref@jnl{Planet.~Space~Sci.}}
\def\procspie{\jref@jnl{Proc.~SPIE}}

\def\ion#1#2{#1$\;${\small\rm\@Roman{#2}}\relax}

\newcount\lecurrentfam
\def\LaTeX{\lecurrentfam=\the\fam \leavevmode L\raise.42ex
\hbox{$\fam\lecurrentfam\scriptstyle\kern-.3em A$}\kern-.15em\TeX}
\def\sizrpt{
(\fontname\the\font): em=\the\fontdimen6\font, ex=\the\fontdimen5\font
\typeout{
(\fontname\the\font): em=\the\fontdimen6\font, ex=\the\fontdimen5\font
}}
\definecolor{dark}{rgb}{ 0.34, 0.09, 0.49}
\def\JWST{{\it JWST}}

\usepackage{fancyhdr,url}
\usepackage{transparent}
\usepackage{tablefootnote}
\usepackage{mathpazo}

\begin{document}

\noindent
{\Huge\bfseries
Galaxies as Stochastic Systems \\[0.5em]
\Large Why the next breakthrough in galaxy evolution requires one hundred million spectra}

\vfill
\thispagestyle{empty}
\noindent
\vspace{1em}
\begin{flushleft}
\textbf{Authors:}
\vspace{1em}

\begin{tabular}{p{0.2cm} l p{1cm} l}
& \multicolumn{3}{p{14cm}}{\it
\textbf{Sandro Tacchella}$^1$  \textit{(Cavendish Laboratory, University of Cambridge, UK)},
\textbf{Vasily Belokurov}  \textit{(Institute of Astronomy, University of Cambridge, UK)},
\textbf{Harry T.~J. Bevins} \textit{(Cavendish Laboratory, University of Cambridge, UK)}, 
\textbf{Roberto Maiolino} \textit{(Cavendish Laboratory, University of Cambridge, UK)}, 
\textbf{Hiranya V. Peiris} \textit{(Institute of Astronomy, University of Cambridge, UK)}, 
\textbf{Lucia Pozzetti} \textit{(INAF Bologna, IT)}, 
\textbf{Mark T. Sargent} \textit{(Laboratoire d’Astrophysique, EPFL, CH)}}
\end{tabular}
\end{flushleft}
\footnotetext[1]{\url{st578@cam.ac.uk}}

\vspace{2em}
\vfill

{
\begin{flushleft}
\end{flushleft}
}

\newpage{}

\vspace{-0.35cm}
\begin{tcolorbox}[boxsep=3pt,left=0pt,right=0pt,top=1pt,bottom=1pt,colframe=boxcolor,colback=boxcolor]
\justify
\textbf{Summary:} 
Each galaxy is observed only once along its life, making galaxy evolution fundamentally an inverse statistical problem: time-dependent physics must be inferred from ensembles of single-epoch snapshots. To move beyond descriptive scaling relations toward physical regulation mechanisms of star formation, quenching, chemical enrichment and black hole growth, galaxies must be treated as realizations of a stochastic process whose hyper-parameters (e.g., correlation timescales, burstiness, duty cycles) are inferred hierarchically. This demands both depth and scale: continuum S/N sufficient for absorption-line ages and chemistry, and samples far larger than those in SDSS, DESI, 4MOST or MOONS, which provide either depth or size but not both across $0<z<3$. Once the relevant axes of mass, redshift, environment, structure and evolutionary phase are populated, the requirement naturally rises from $10^7$ to $\sim10^8$ galaxies. This is the regime where stochastic hyper-parameters can be well constrained and where comparisons to simulations and cosmological forward models become limited by theory rather than observations. We outline the science enabled by such a programme and the corresponding requirements for a future ESO wide-field spectroscopic facility capable of delivering tens to hundreds of millions of rest-UV–optical spectra over $0\lesssim z\lesssim3$.
\end{tcolorbox}

\vspace{-0.5cm}
\section{Scientific Context and Open Questions}\label{Sec1}
\vspace{-0.3cm}

Galaxies are not static containers of stars and gas; they are time-variable, self-regulating systems in which gas accretion, star formation, feedback and environmental processes govern the baryon cycle \cite{lilly13_bathtub, tacchella16_MS}. Yet we never see the same galaxy twice: each object is a single snapshot of a long and complex history. Cosmic evolution must therefore be reconstructed statistically, by comparing ensembles of galaxies across mass, redshift and environment \cite{abramson16}. Archaeological reconstructions are feasible for a few nearby systems, but suffer from degeneracies even with high-quality spectra, as the mixing of \emph{in situ} and \emph{ex situ} stars through mergers is a fundamental limitation \cite{tacchella22_quench, cochrane25}.

Over the past two decades, imaging and spectroscopic surveys have established a coherent empirical framework for galaxy evolution \cite{forster-schreiber20}: the star-forming main sequence, the growth of quiescence, the mass–metallicity relation and environmental trends. SDSS and GAMA provide statistical power at $z<0.3$; LEGA-C, DEEP2, zCOSMOS, VIPERS, VUDS and VANDELS extend this to $z\sim4$ with limited samples. Ongoing programmes such as DESI, 4MOST, PFS and MOONS will deliver millions of spectra, but with insufficient depth for stellar-population work for the majority of galaxies. These surveys capture time evolution only in compressed form, mean trends and scatter, rather than the temporal structure of star formation or the frequency of rare events.

Theoretically, star formation is viewed as a stochastic, self-regulated process shaped by gas inflow, gravitational instability, stellar and black-hole feedback, and environment \cite{naab17}. This motivates population-level quantities such as the temporal power spectral density (PSD) of star-formation variability, correlation timescales, and duty cycles for bursts, quenching and rejuvenation \cite{kelson14, forbes14b, caplar19, tacchella20}. Recent \JWST\ observations have revealed pronounced burstiness in early, low-mass galaxies \cite{tacchella23_metal, endsley24, simmonds25}. Different physical models predict distinct PSD shapes \cite{iyer20}, but current data provide only weak constraints because the combined required \emph{depth} (continuum S/N $\sim$ 20 per Å) and \emph{sample size}.

The emerging realisation is that the key observables of galaxy evolution in the 2040s will not be individual stellar masses or SFRs, but the \emph{hyper-parameters} of the stochastic processes that generate them. Measuring these requires (i) spectroscopic diagnostics that jointly probe gas conditions and fossil stellar populations across the rest-frame UV–optical, and (ii) samples large enough to map high-dimensional distributions, including rare, short-duty-cycle phases at the $\sim0.01\%$ level. Achieving several thousand objects per bin across mass, redshift and environment pushes beyond current $10^5$–$10^6$ surveys to the $\sim10^7$–$10^8$ regime required to characterise stochasticity. This spectroscopic layer will integrate with upcoming wide-field imaging (Euclid, Roman, LSST), which together provide the multi-wavelength context for interpreting these stochastic hyper-parameters.

\vspace{-0.3cm}
\section{Science Case for the 2040s}\label{Sec2}
\vspace{-0.2cm}

\subsection*{From scaling relations to stochastic laws}

The limitation of current surveys is not only sample size, but the statistical framework in which galaxy data are interpreted. For any individual galaxy, its spectrum and photometry provide noisy, highly degenerate constraints on quantities such as its star-formation history, chemical enrichment and dust geometry. Modern spectral--photometric inference can treat star-formation histories (SFHs) as stochastic processes with physically motivated temporal priors and embed them in generative population models \cite{thorp24, wan24}. In this regime, single galaxies yield only weak information about burstiness, correlation timescales or burst cycles; the constraining power instead comes from the \emph{joint distribution} of millions of objects, accessed through hierarchical Bayesian inference.

This perspective has two consequences. First, galaxy evolution becomes statistically analogous to cosmology: the goal shifts from fitting each object to inferring the hyper-parameters of a generative model describing the full population and its intrinsic scatter. Second, required sample sizes are no longer set by populating a few scaling relations, but by constraining a genuinely high-dimensional hyper-parameter space while marginalising over noise, selection and modelling systematics. In this regime, surveys of order $10^7$ galaxies are the natural scale needed.

Once the key axes of redshift, stellar mass, environment (density and cosmic-web position), morphology (size, compactness, bulge fraction) and evolutionary phase (star-forming, transitioning, quiescent, starburst, rejuvenated, or AGN) are considered, even coarse binning yields hundreds of distinct cells. Recovering the shapes of the underlying distributions in each cell, rather than just their means and scatters, requires samples of order $\sim10^3$ galaxies per bin. Measuring temporal PSD shapes, correlation timescales and the incidence of rare phases with $0.1$–1\% duty cycles makes the requirement more stringent: obtaining several thousand objects in each state across mass and environment naturally drives us to $\gtrsim10^7$ spectra. Thus the million-spectra scale is not a slogan but the threshold where galaxy evolution becomes a problem in stochastic population physics. Reaching tens of million spectra allows clear discrimination between competing models of star-formation regulation and quenching, rather than merely fitting parametric trends.

\subsection*{Key questions enabled by a hundred-million–spectra programme}

A wide-field spectroscopic survey delivering $\sim10^8$ galaxy spectra over $0 \lesssim z \lesssim 3$ would open an entirely new regime in galaxy evolution: the ability to measure population-level stochastic laws, rare phases, and environmental dependencies with cosmology-grade precision.

\textit{How does the temporal structure of star formation depend on stellar mass, environment and cosmic time?}  
Joint emission and absorption line diagnostics allow the variance of SFHs on different timescales to be inferred, testing whether high-mass galaxies have less bursty histories than low-mass ones, and how environment modulate variability and long-timescale regulation.

\textit{What are the duty cycles and timescales of quenching and rejuvenation?}  
Rather than labelling galaxies as ``star-forming'' or ``quiescent'', a $10^8$-object sample allows the probability flux through the SFR–$M_\star$ plane to be measured as a function of environment and structure, constraining the prevalence of rapid quenching and the frequency of rejuvenation episodes.

\textit{How does environment regulate SFHs across the cosmic web?}  
Sub-Mpc mapping of the cosmic web at $z\sim0.5$–3, coupled to deep spectroscopy, enables tests of whether the star-formation activity is suppressed or enhanced near filaments and nodes, how rapidly satellites transform after infall, and whether large-scale conformity persists on multi-Mpc scales to high redshifts.

\textit{What are the multi-dimensional distributions of metallicity and abundance ratios?}  
High-S/N absorption line spectroscopy across $10^8$ galaxies permits hierarchical inference of stellar metallicity and $\alpha$-enhancement distributions at fixed mass, SFR and environment. These chemical clocks constrain star-formation timescales, allow comparison with stellar-population ages, and separate \emph{in situ} from \emph{ex situ} mass assembly.

\textit{How do empirical stochastic laws map onto simulations?}  
Next-generation simulations will predict full stochastic hyper-parameters (PSD shapes, correlation lengths, duty cycles, environmental response) for millions of galaxies. A hundred-million–spectra survey provides the first direct observational inference of these quantities, enabling comparisons at the level of stochastic laws rather than mean trends.

\textit{How does empirical galaxy physics calibrate cosmological forward models?}  
Generative models of the galaxy population are already transforming the forward-modelling needed for accurate cosmological inference \cite{thorp25}. Deep spectroscopy at the $10^8$-galaxy scale would extend this framework into an entirely new regime, where the model encodes not just scaling relations but the temporal structure of star formation. At this scale, the stochastic laws of galaxy evolution become directly measurable, and the astrophysical calibration needed for precision large-scale structure and weak lensing cosmology moves from aspiration to reality.

\vspace{-0.3cm}
\section{Facility and Technical Requirements}\label{Sec3}
\vspace{-0.2cm}

Achieving these goals requires a next-generation wide-field spectroscopic facility capable of delivering joint emission  and absorption line diagnostics over the rest-frame UV–optical window for $\sim10^7$–$10^8$ galaxies. Here we summarise the key requirements.

\textbf{Spectral range and diagnostics:} At $z\sim0.5$–1.5, the observed optical window provides the core rest-frame optical emission lines (\oii, H$\beta$, \oiii, H$\alpha$, \nii, \sii) alongside key stellar absorption features (Balmer series, 4000~\AA\ break, Ca H\&K, Mg\,b, Fe), enabling constraints on instantaneous gas properties and Gyr-scale SFHs. Extending coverage to $\sim2.2~\mu$m ensures access to rest-UV indices tracing young populations at $z\sim1$ and preserves H$\alpha$ and the full optical suite out to $z\sim3$. A resolving power of $R\sim3000$–6000 and continuum S/N $\sim20$ per \AA\ suffice for recovering stellar ages and abundances, including hierarchical analyses.

\textbf{Survey design and multiplex:} Reaching $\sim10^7$–$10^8$ galaxies requires surveys over $\sim1000$~deg$^2$ with target densities of a few$\times10^4$~deg$^{-2}$. A highly multiplexed MOS with $10^4$–$3\times10^4$ simultaneous fibres or slits over a $\sim2$–3~deg$^2$ field of view would deliver several $\times10^5$ spectra per dark night and reach the required sample in a realistic multi-year programme.

\textbf{Synergies and data infrastructure:} The scientific return relies on complementary imaging from Euclid, Roman, LSST and future UV/IR facilities, as well as cold-gas and high-energy constraints from SKA, ngVLA, ALMA and next-generation X-ray missions. An ESO-led spectroscopic layer provides the dynamical, chemical and temporal information needed to interpret these data, and must be paired with population-level inference tools delivering homogenised measurements and posterior summaries.

\vspace{0.1cm}

\footnotesize{
\noindent
\bibliographystyle{naturemag_notitle}

\begin{thebibliography}{10}
\expandafter\ifx\csname url\endcsname\relax
  \def\url#1{\texttt{#1}}\fi
\expandafter\ifx\csname urlprefix\endcsname\relax\def\urlprefix{URL }\fi
\providecommand{\bibinfo}[2]{#2}
\providecommand{\eprint}[2][]{\url{#2}}

\bibitem{lilly13_bathtub}
\bibinfo{author}{{Lilly}, S.~J.}, \bibinfo{author}{{Carollo}, C.~M.},
  \bibinfo{author}{{Pipino}, A.}, \bibinfo{author}{{Renzini}, A.} \&
  \bibinfo{author}{{Peng}, Y.}
\newblock \emph{\bibinfo{journal}{\apj}} \textbf{\bibinfo{volume}{772}},
  \bibinfo{pages}{119} (\bibinfo{year}{2013}).

\bibitem{tacchella16_MS}
\bibinfo{author}{{Tacchella}, S.} \emph{et~al.}
\newblock \emph{\bibinfo{journal}{\mnras}} \textbf{\bibinfo{volume}{457}},
  \bibinfo{pages}{2790--2813} (\bibinfo{year}{2016}).

\bibitem{abramson16}
\bibinfo{author}{{Abramson}, L.~E.} \emph{et~al.}
\newblock \emph{\bibinfo{journal}{\apj}} \textbf{\bibinfo{volume}{832}},
  \bibinfo{pages}{7} (\bibinfo{year}{2016}).

\bibitem{tacchella22_quench}
\bibinfo{author}{{Tacchella}, S.} \emph{et~al.}
\newblock \emph{\bibinfo{journal}{\apj}} \textbf{\bibinfo{volume}{926}},
  \bibinfo{pages}{134} (\bibinfo{year}{2022}).

\bibitem{cochrane25}
\bibinfo{author}{{Cochrane}, R.~K.}
\newblock \emph{\bibinfo{journal}{\mnras}} \textbf{\bibinfo{volume}{544}},
  \bibinfo{pages}{1530--1540} (\bibinfo{year}{2025}).

\bibitem{forster-schreiber20}
\bibinfo{author}{{F{\"o}rster Schreiber}, N.~M.} \& \bibinfo{author}{{Wuyts},
  S.}
\newblock \emph{\bibinfo{journal}{\araa}} \textbf{\bibinfo{volume}{58}},
  \bibinfo{pages}{661--725} (\bibinfo{year}{2020}).

\bibitem{naab17}
\bibinfo{author}{{Naab}, T.} \& \bibinfo{author}{{Ostriker}, J.~P.}
\newblock \emph{\bibinfo{journal}{\araa}} \textbf{\bibinfo{volume}{55}},
  \bibinfo{pages}{59--109} (\bibinfo{year}{2017}).

\bibitem{kelson14}
\bibinfo{author}{{Kelson}, D.~D.}
\newblock \emph{\bibinfo{journal}{ArXiv e-prints}}  (\bibinfo{year}{2014}).

\bibitem{forbes14b}
\bibinfo{author}{{Forbes}, J.~C.}, \bibinfo{author}{{Krumholz}, M.~R.},
  \bibinfo{author}{{Burkert}, A.} \& \bibinfo{author}{{Dekel}, A.}
\newblock \emph{\bibinfo{journal}{\mnras}} \textbf{\bibinfo{volume}{443}},
  \bibinfo{pages}{168--185} (\bibinfo{year}{2014}).

\bibitem{caplar19}
\bibinfo{author}{{Caplar}, N.} \& \bibinfo{author}{{Tacchella}, S.}
\newblock \emph{\bibinfo{journal}{\mnras}} \textbf{\bibinfo{volume}{487}},
  \bibinfo{pages}{3845--3869} (\bibinfo{year}{2019}).

\bibitem{tacchella20}
\bibinfo{author}{{Tacchella}, S.}, \bibinfo{author}{{Forbes}, J.~C.} \&
  \bibinfo{author}{{Caplar}, N.}
\newblock \emph{\bibinfo{journal}{\mnras}} \textbf{\bibinfo{volume}{497}},
  \bibinfo{pages}{698--725} (\bibinfo{year}{2020}).

\bibitem{tacchella23_metal}
\bibinfo{author}{{Tacchella}, S.} \emph{et~al.}
\newblock \emph{\bibinfo{journal}{\mnras}} \textbf{\bibinfo{volume}{522}},
  \bibinfo{pages}{6236--6249} (\bibinfo{year}{2023}).

\bibitem{endsley24}
\bibinfo{author}{{Endsley}, R.} \emph{et~al.}
\newblock \emph{\bibinfo{journal}{\mnras}} \textbf{\bibinfo{volume}{533}},
  \bibinfo{pages}{1111--1142} (\bibinfo{year}{2024}).

\bibitem{simmonds25}
\bibinfo{author}{{Simmonds}, C.} \emph{et~al.}
\newblock \emph{\bibinfo{journal}{\mnras}}  (\bibinfo{year}{2025}).

\bibitem{iyer20}
\bibinfo{author}{{Iyer}, K.~G.} \emph{et~al.}
\newblock \emph{\bibinfo{journal}{\mnras}} \textbf{\bibinfo{volume}{498}},
  \bibinfo{pages}{430--463} (\bibinfo{year}{2020}).

\bibitem{thorp24}
\bibinfo{author}{{Thorp}, S.} \emph{et~al.}
\newblock \emph{\bibinfo{journal}{\apj}} \textbf{\bibinfo{volume}{975}},
  \bibinfo{pages}{145} (\bibinfo{year}{2024}).

\bibitem{wan24}
\bibinfo{author}{{Wan}, J.~T.} \emph{et~al.}
\newblock \emph{\bibinfo{journal}{\mnras}} \textbf{\bibinfo{volume}{532}},
  \bibinfo{pages}{4002--4025} (\bibinfo{year}{2024}).

\bibitem{thorp25}
\bibinfo{author}{{Thorp}, S.} \emph{et~al.}
\newblock \emph{\bibinfo{journal}{\apj}} \textbf{\bibinfo{volume}{993}},
  \bibinfo{pages}{240} (\bibinfo{year}{2025}).

\end{thebibliography}

}

\end{document}